\documentclass{elsart}

\usepackage{adnd}
\usepackage{longtable}

\usepackage{mathptmx}

\usepackage{amsmath}

\usepackage{amssymb,amstext}
\usepackage[pdftex,
        letterpaper=true,
        hyperindex=true,
        breaklinks=true,
        colorlinks=false,
        citecolor=blue,
        pdftitle={},
        pdfauthor={}]
{hyperref}
\usepackage{graphicx}

\begin{document}

\begin{frontmatter}

\journal{Atomic Data and Nuclear Data Tables}

\copyrightholder{Elsevier Science}

\runtitle{Gold}
\runauthor{Schuh}

%% Author, fill in article title here

\title{Discovery of the Gold Isotopes}

%% Fill in author list here

\author{A. Schuh},
\author{A. Fritsch},
\author{J.Q. Ginepro},
\author{M. Heim},
\author{A. Shore},
\and
\author{M.~Thoennessen\corauthref{cor}}\corauth[cor]{Corresponding author.}\ead{thoennessen@nscl.msu.edu}

\address{National Superconducting Cyclotron Laboratory and \\ Department of Physics and Astronomy, Michigan State University, \\East Lansing, MI 48824, USA}

\date{26.12.2008} %please do not use \today, use actual date of version

\begin{abstract}
Thirty-six gold isotopes have so far been observed; the discovery of these isotopes is discussed.  For each isotope a brief summary of the first refereed publication, including the production and identification method, is presented.
\end{abstract}

\end{frontmatter}

%%% Keywords and subject classification are not used in ADNDT
%%%\begin{keywords}
%%%Insert list of keywords here.
%%%\end{keywords}

%%%\begin{subject}[Insert header for classifications]
%%%Use only if your journal has a subject classification requirement
%%%\end{subject}

%%% The table of contents should start a new page. This command will
%%% automatically pull all the unstarred \section, \subsection and
%%% \subsubsection titles into the Contents. Starred versions need to be
%%% done manually using
%%%            \addcontentsline{toc}{[[sub]sub]section}{Section title}
%%% at the correct place. Examples are given below.

%%% The lists of data figures and data tables are created automatically
%%% by the \listofDfigures and \listofDtables commands.

\newpage
\tableofcontents
%%\listofDfigures
\listofDtables

\vskip5pc

\section{Introduction}\label{s:intro}

In this third paper in the series of the discovery of isotopes \cite{Gin09,Sch09} the discovery of the gold isotopes is discussed. Previously, the discovery of cerium \cite{Gin09} and arsenic \cite{Sch09} isotopes was discussed. The purpose of this series is to document and summarize the discovery of the isotopes. Guidelines for assigning credit for discovery are (1) clear identification, either through decay-curves and relationships to other known isotopes, particle or $\gamma$-ray spectra, or unique mass and Z-identification, and (2) publication of the discovery in a refereed journal. The authors and year of the first publication, the laboratory where the isotopes were produced as well as the production and identification methods are discussed. When appropriate, references to conference proceedings, internal reports, and theses are included. When a discovery includes a half-life measurement the measured value is compared to the currently adapted value taken from the NUBASE evaluation \cite{Aud03} which is based on ENSDF database \cite{ENS08}.

\section{Discovery of $^{170-205}$Au}

Thirty-six gold isotopes from A = $170-205$ have been discovered so far; these include 1 stable, 27 proton-rich and 8 neutron-rich isotopes.  According to the HFB-14 model \cite{Gor07}, $^{262}$Au should be the last even particle stable neutron-rich nucleus and the odd particle stable neutron-rich nuclei should continue through $^{265}$Au. Thus, there remain 60 neutron-rich isotopes to be discovered. The proton dripline has been reached and several isotopes beyond the dripline with long lifetimes have been observed. It is estimated that 6 additional nuclei beyond the proton dripline could live long enough to be observed \cite{Tho04}. Less than 40\% of all possible gold isotopes have been produced and identified so far.

\begin{figure}
	\centering
	\includegraphics[scale=.5]{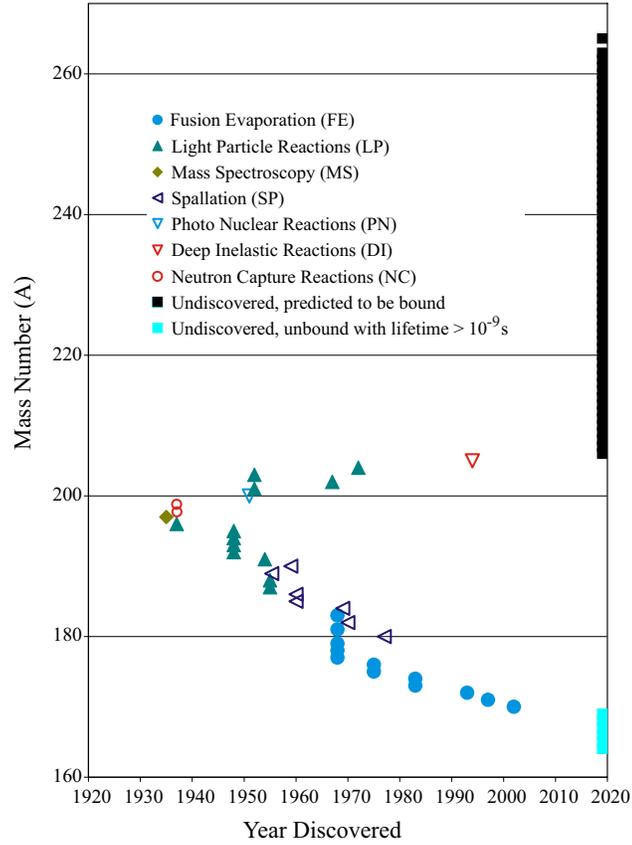}
	\caption{Gold isotopes as a function of time they were discovered. The different production methods are indicated. The solid black squares on the right hand side of the plot are isotopes predicted to be bound by the HFB-14 model.  On the proton-rich side the light blue squares correspond to unbound isotopes predicted to have lifetimes larger than $\sim 10^{-9}$~s.}
\label{f:year}
\end{figure}

Figure \ref{f:year} summarizes the year of first discovery for all gold isotopes identified by the method of discovery.  The range of isotopes predicted to exist is indicated on the right side of the figure.  The radioactive gold isotopes were produced using fusion evaporation (FE), deep-inelastic reactions (DI), light-particle reactions (LP), photo-nuclear (PN), neutron-capture (NC) and spallation reactions (SP). Heavy ions are all nuclei with an atomic mass larger than A=4 \cite{Gru77}. Light particles also include neutrons produced by accelerators. In the following, the discovery of each gold isotope is discussed in detail.

\subsection*{$^{170}$Au}\vspace{-0.85cm}

\textit{New results on proton emission from odd-odd nuclei} reported the discovery of $^{170}$Au by Mahmud \textit{et al.} using the ATLAS accelerator at Argonne National Laboratory in 2002 \cite{Mah02}.  The isotope was produced in the fusion-evaporation reaction $^{96}$Ru($^{78}$Kr,1p3n) at a beam energy of 400 MeV and identified by its proton emission: ``The proton decay of $^{170}$Au is clearly present at an energy E$_p$ = 1735(9)keV ... with a half-life of 570$^{+350}_{-150} \mu s$.'' This corresponds to an isomeric state of $^{170}$Au.

\subsection*{$^{171}$Au}\vspace{-0.85cm}

Davids \textit{et al.} first observed $^{171}$Au at Argonne National Laboratory in 1997 \cite{Dav97} reported in \textit{New proton radioactivities $^{165, 166, 167}$Ir and $^{171}$Au}. The fusion-evaporation reaction $^{96}$Ru($^{78}$Kr, \textit{p}2\textit{n}) at a beam energy of 397 MeV was used. ``The proton emitters were each identified by position, time, and energy correlations between the implantation of a residual nucleus into a double-sided silicon strip detector, the observation of a decay proton, and the subsequent observation of a decay alpha particle from the daughter nucleus.'' The half-life was determined to be 1.02(15) ms and corresponds to an isomeric state.

\subsection*{$^{172}$Au}\vspace{-0.85cm}

$^{172}$Au was discovered by Sellin \textit{et al.} at Daresbury, England in 1993, as reported in \textit{Alpha decay of the new isotope $^{172}$Au} \cite{Sel93}.  The fusion evaporation reaction $^{106}$Cd($^{70}$Ge,1p3n) at 354 MeV was used. $^{172}$Au was identified by its $\alpha$-decay following the implantation of residues at the end of the Daresbury recoil mass separator.  They observed ``a new alpha decay line at an energy above that of the lightest known gold isotope $^{173}$Au.  This activity occurs within the A = 172 region of the strip detector and is assigned to $^{172}$Au on the basis of alpha decay Q-value systematics.''  From the 43 observed counts a half-life of 4(1) ms was extracted for the $\alpha$-decay, which is included in the determination of the currently accepted value of 4.7(1.1) ms.

\subsection*{$^{173-174}$Au}\vspace{-0.85cm}

\textit{Alpha Decay of New Neutron Deficient Gold, Mercury and Thallium Isotopes} reported the first discovery of $^{173}$Au and $^{174}$Au by Schneider \textit{et al.} in 1983 at the Gesellschaft f\"{u}r Schwerionenforschung (GSI) in Germany \cite{Sch83}. The isotopes were produced in fusion evaporation reactions with a $^{92}$Mo beam of energies between 4.5 A$\cdot$MeV and 5.4 A$\cdot$MeV and separated with the velocity filter SHIP. The $\alpha$-decay of $^{173}$Au ``could be correlated to the known $\alpha$-decays of their daughters'' with a halflife of 59$^{+45}_{-18}$ ms. The assignment of $^{174}$Au was made because ``this activity was seen to peak at excitation energies (43.5 MeV) intermediate between those of the unambiguously identified isotopes $^{173}$Au and $^{175}$Au, and the corresponding decay energy does not match with any of the previously known lower Z (Pt, Ir) isotopes that were energetically possible in the particular region.'' The half-life of $^{174}$Au was measured to be 120(20) ms. Both half-lives are close to the adapted values of 25(1) ms ($^{173}$Au) and 139(3) ms ($^{174}$Au).

\subsection*{$^{175-176}$Au}\vspace{-0.85cm}

Cabot \textit{et al.} reported the first discovery of $^{175}$Au and $^{176}$Au in 1975 at Orsay, France \cite{Cab75}. The paper \textit{Ca induced reactions on $^{141}$Pr and $^{150}$Sm: New gold and lead isotopes $^{176}$Au, $^{175}$Au, $^{185}$Pb} describes excitation function of fusion evaporation reactions and $\alpha$-decay measurements. $^{175}$Au and $^{176}$Au were produced in the reactions $^{141}$Pr($^{40}$Ca,6n) and $^{141}$Pr($^{40}$Ca,5n), respectively. ``Their shifts of $\approx$ 12 MeV excitation energy with reference to the $^{177}$Au curves lead us to assign these transitions to the $\alpha$-decay of $^{176}$Au produced by ($^{40}$Ca,5n). Similarly the behavior of the yield of the 6.44 MeV peak indicates $^{175}$Au as the isotope associated with it.'' For $^{176}$Au a half-life of 1.25(30) s was deduced. This value is consistent with the adapted value of 1.08(17) s, however, the energy of the $\alpha$-decay could not be confirmed.

\subsection*{$^{177-179}$Au}\vspace{-0.85cm}

$^{177}$Au, $^{178}$Au and $^{179}$Au were first observed by Siivola at Lawrence Radiation Laboratory at the University of California, Berkeley, in 1968 \cite{Sii68}. The article \textit{Alpha-Active Gold Isotopes} reported that the isotopes were seen when $^{168}$Yb was bombarded with $^{18}$F.  The three new isotopes were created following the evaporation of 9, 8 and 7 neutrons. ``The 5.848 MeV activity can be assigned to $^{179}$Au because in the $^{20}$Ne + $^{169}$Tm bombardments it is produced in large quantities at 130 MeV excitation, which is above the expected position for a ($^{20}$Ne, 9n) reaction.''  The half-lives of $^{177}$Au, $^{178}$Au and $^{179}$Au were determined to be 1.3(4)~s, 2.6(5)~s and 7.2(5)~s respectively; this is the only half-life measurement for $^{178}$Au, the value for $^{177}$Au is consistent with the adapted values of 1.46(3)~s and the half-life of  $^{179}$Au is included in the current average value of 7.1(3)~s. It should be mentioned that $^{179}$Au was first reported by A.G. Demin \textit{et al.} \cite{Dem68}. Their manuscript was received on June 20, 1967 while Siivola's paper was received only on October 24, 1967. However, Demin had access to the unpublished report by Siivola \cite{Sii65} and frequently refered to this work.

\subsection*{$^{180}$Au}\vspace{-0.85cm}

\textit{On-line $\gamma$-ray spectroscopic investigation of the $^{180}$Hg(T$_{1/2}$ = 3s) Decay Chain} reported the discovery of $^{180}$Au by Husson \textit{et al.} in 1977 at CERN \cite{Hus77}.  $^{180}$Au was produced following the $\beta$-decay of $^{180}$Hg produced in the (p,3pxn) spallation reaction from a 600 MeV proton beam on a lead target. The identification was achieved with $\beta-\gamma$ coincidence measurements of the separated $^{180}$Hg ions.  ``...no ambiguity occurs for the $\gamma$ lines attributed to the $^{180}$Hg $\rightarrow$ $^{180}$Au decay.'' The half-life was determined to be 8.1(3)s which is currently the only half-life measurement for this isotope.

\subsection*{$^{181}$Au}\vspace{-0.85cm}

Siivola discovered $^{181}$Au during the same experiment as $^{177}$Au, $^{178}$Au and $^{179}$Au in 1968 at the University of California-Berkeley using a $^{19}$F beam on a $^{168}$Yb target and was also discussed in \textit{Alpha-Active Gold Isotopes} \cite{Sii68}.  Studying the alpha spectra ``gives the mass numbers of the 5.073, 5.343, 5.482 and 5.623 MeV alpha groups as 185, 183 and 181, respectively. The isotope $^{181}$Au appears to have two alpha groups.'' The half-life was determined to be 11.5(1) s which is included in the currently adapted value of 13.7(14) s.

\subsection*{$^{182}$Au}\vspace{-0.85cm}

$^{182}$Au was discovered by Hansen \textit{et al.} in 1970 with the ISOLDE facility at CERN \cite{Han70}.  The data reported in \textit{Studies of the $\alpha$-active isotopes of mercury, gold and platinum} for $^{178}$Pt were produced following spallation reactions of 600 MeV protons on a molten lead target. The isotope was identified from the growth and decay spectrum of high-energy positrons from the parent nucleus $^{182}$Hg: ``We have performed a half-life measurement using the high-energy positrons ... and found 19 $\pm$ 2 s.'' This half-life is close to the adapted value of 15.5(4)s.

\subsection*{$^{183}$Au}\vspace{-0.85cm}

$^{183}$Au was first seen by Siivola at the University of California, Berkeley, in 1968 during the same experiment as $^{177-179}$Au and $^{181}$Au through bombardment of $^{168}$Yb with $^{19}$F ions \cite{Sii68}.  \textit{Alpha-Active Gold Isotopes} discussed the alpha spectra which gave ``the mass numbers of the 5.073, 5.343, 5.482 and 5.623 MeV alpha groups as 185, 183 and 181, respectively.'' The half-life was determined to be 45(4) s and is included in the adapted value of 42.8(10) s.

\subsection*{$^{184}$Au}\vspace{-0.85cm}

Hansen \textit{et al.} reported the first observation of $^{184}$Au in the paper \textit{Decay Characteristics of Short-Lived Radio-Nuclides Studied by On-Line Isotope Separator Techniques} in 1969 \cite{Han69}. 600 MeV protons from the CERN synchrocyclotron bombarded a lead target and mercury was separated using the ISOLDE facility. Electron capture, $\beta$- and $\gamma$-rays were measured to identify $^{182}$Au. The paper summarized the ISOLDE program and did not contain details about the individual nuclei other than in tabular form. The detailed analysis was published in reference \cite{Fin72}. The measured half-life of 47(3)~s corresponds to an isomeric state.

\subsection*{$^{185-186}$Au}\vspace{-0.85cm}

$^{185}$Au and $^{186}$Au were first observed by Albouy \textit{et al.} in 1960 with spallation reactions: \textit{Noveaux isotopes de p\'eriode courte obtenus par spallation de l'or} \cite{Alb60}. The gold targets were bombared with 155 MeV protons from the Orsay synchro-cyclotron. ``L'intensit\'e obtenue pour les masses 187, 186 et 185 rend peu pr\'ecise la d\'etermination des \'energies des raies $\gamma$ correspondant aux masss 187 et 186 et ne nous a pas permis d'indentifier des raies $\gamma$ pour la cha\^ine de masses 185.'' (The intensity obtained for the masses 187, 186 and 185 makes a precise determination of the $\gamma$-ray energies corresponding to masses 187 and 186 possible but has not allowed us to identify $\gamma$-rays for the mass of 185.) The quoted half-life of $^{185}$Au (7 m) corresponds to an isomeric state and the half-life of 12~m for $^{186}$Au agrees with the accepted value of 10.7(5)m. It should be mentioned that the decay of the ground state of $^{185}$Au (4.3(2) m) had been observed previously, however, the mass could only tentatively be assigned between 184 and 187 \cite{Ras53}.

\subsection*{$^{187-188}$Au}\vspace{-0.85cm}

Smith and Hollander first observed $^{187}$Au and $^{188}$Au at the Berkeley Radiation Laboratory in 1955 \cite{Smi55} and the results were reported in \textit{Radiochemical Study of Neutron-Deficient Chains in the Noble Metal Region}. The isotopes were primarily produced in (p,xn) reactions with protons of energies between 50 and 130 MeV accelerated with the 184-cyclotron. Additional measurements were performed with 32 MeV protons from the Berkeley linear accelerator and protons and heavy ions from the 60-inch cyclotron. Identification was achieved with timed chemical separation. Characteristic $\gamma$-ray spectra were measured with a NaI detector. The half-life of $^{187}$Au was extracted as $\sim$15~m which is close to the currently accepted value of 8.4(3)~m. ``In two 130-MeV proton bombardments where platinum was milked from a gold parent fraction, we have obtained preliminary evidence that the half-life of Au$^{188}$ is of the order of 10 minutes.'' The half-life is consistent with the present value of 8.84(6)~m.

\subsection*{$^{189}$Au}\vspace{-0.85cm}

N. Poffe \textit{et al.} reported the first observation of $^{189}$Au in \textit{R\'eactions (p,xn) induites dans l'or par des protons de 155 MeV} in 1960 \cite{Pof60}. The target was bombarded with 155 MeV protons from the synchrocyclotron of the Paris Faculty of Sciences and $^{189}$Au was identified by half-life and $\gamma$-ray measurements following magnetic separation: ``Half-lives and main $\gamma$-ray energies have been measured for $^{190}$Hg, $^{189}$Hg, $^{188}$Hg and their daugther products.'' The quoted half-life of 30 m agrees with the currently accepted value of 28.7(3) m. Previous claims of the observation of $^{189}$Au with a half-life of about 42 m \cite{Smi55,Cha57} was most likely due to the decay of $^{190}$Au.

\subsection*{$^{190}$Au}\vspace{-0.85cm}

G. Albouy \textit{et al.} observed $^{190}$Au for the first time in 1959 at Orsay: \textit{Isotopes de nombe de masse 190 du mercure et de l'or} \cite{Alb59}. Mercury isotopes were produced in (p,xn) spallation reactions and mass 190 was selected with an isotope separator. K-X-rays were detected with a NaI(Tl) detector. ``Nous avons observ\'e dans la d\'ecroissance du rayonnement K principalement deux p\'eriodes en filiation 21$\pm$2 mn suivie de 45$\pm$3 mn. Nous attribuons ces p\'eriodes respectivement \'a $^{190}$Hg et \'a son descendant $^{190}$Au.'' (We observed the decay by K-radiation to consist mainly of two correlated periods, 21$\pm$2 m followed by 45$\pm$3 m. We attributed these periods with $^{190}$Hg and its daughter $^{190}$Au, respectively.) The half-life for $^{190}$Au agrees with the accepted value of 42.8(10) m.

\subsection*{$^{191}$Au}\vspace{-0.85cm}

$^{191}$Au was first clearly identified by Gillon \textit{et al.} from Princeton University in \textit{Nuclear Spectroscopy of Neutron-Deficient Hg Isotopes} in 1954 \cite{Gil54}. The experiment was performed at the Harvard Cyclotron Laboratory using the reaction $^{197}$Au(p,7n) at 65 MeV. Conversion electron spectra were measured with a 119-gauss magnet and K-X-ray lines were attributed to $^{191}$Au following the decay of $^{191}$Hg. In addition ``It is likely that the 252.8-kev $\gamma$ ray is the \textit{d$_{5/2}-$d$_{3/2}$} transition in Au$^{191}$.'' The estimated half-life of 4 h agrees with the adapted value of 3.18(8) h. The previous provisional assignment of a one day half-life in $^{191}$Au \cite{Wil48} deviates significantly from this value. In addition, the paper stated that ``the isotope has not been directly observed.''

\subsection*{$^{192-195}$Au}\vspace{-0.85cm}

$^{192-195}$Au were observed by Wilkinson at the University of California-Berkeley in 1948, as reported in \textit{Some Isotopes of Platinum and Gold} \cite{Wil48}. Helium and deuteron beams of the 60-inch Crocker Laboratory cyclotron bombarded iridium and platinum targets, respectively. Half-lives were measured following chemical separation. For an observed 4.7 hour activity a ``provisional assignment to mass 192 has been made on the basis of reaction yields, and absence of any daughter activity.''  A 15.8 hour half-life was also seen and was assigned to $^{193}$Au, because ``a half-life for Au$^{193}$ longer than 16 hours and of the order of days might have been expected from consideration of the stability of isotopes of elements of odd Z in this region, but since the allocation of Pt$^{193}$ seems reliable, the gold parent has been allocated similarly.''  Because ``chemical separations give no evidence of a platinum daughter activity of half-life greater than a few minutes," a 39.5 hour decay was assigned to $^{194}$Au.  A 190-day activity was observed, as well: ``allocation to mass 195 is consistent with the yields and the half-life is consistent with expectations for a gold isotope of this mass number.''  All of these half-life values are consistent with the currently accepted values.

\subsection*{$^{196}$Au}\vspace{-0.85cm}

McMillan \textit{et al.} from the University of California at Berkeley identified $^{196}$Au for the first time in 1937 in \textit{Neutron-Induced Radioactivity of the Noble Metals} \cite{McM37}. Following the irradiation of a gold target with fast neutrons produced with a deuteron beam on lithium, activities of 13~h and 4-5~d were observed. ``This leads to some difficulty in placing the 13-hr. and 4-5-day periods in the system of isotopes, since only one space (Au$^{196}$) satisfies the necessary conditions for both of them.'' These half-lives are in reasonable agreement with the accepted values of 9.6(1) h and 6.1669(6) d for an isomeric and the ground state of $^{196}$Au, respectively.

\subsection*{$^{197}$Au}\vspace{-0.85cm}

Demster reported the observation of stable $^{197}$Au through mass spectography in 1935 at the University of Chicago \cite{Dem35}.  \textit{Isotopic Constitution of Palladium and Gold} posits that this is the only stable isotope of gold, as the experiment ``failed to show any trace of a heavier isotope.''

\subsection*{$^{198}$Au}\vspace{-0.85cm}

The identification of $^{198}$Au was first reported by Pool \textit{et al.} from the University of Michigan in 1937 in the paper \textit{A Survey of Radioactivity Produced by High Energy Neutron Bombardment} \cite{Poo37}. The bombardment of gold with 20 MeV neutrons resulted in the observation of weak intensities of half-lives of 17 m and 2.5~d which were assigned to $^{198}$Au. The paper presented the results for a large number of elements and ``The assignments of the periods is tentative and is based upon evidence from the sign of the emitted beta-particle, the chemical separations and known periods from other sources.'' The longer half-life agrees with the half-life of the ground state (2.69517(21) d) and an isomeric state (2.27(2) d). It should be mentioned that the paper was received on July 3, 1937, while a paper of McMillan \textit{et al.} was received only about two weeks later, on July 19, 1937, and it assigned the observation of a 2.7 d half-life to $^{198}$Au \cite{McM37}. This activity had actually been observed already by Fermi \textit{et al.} and Amaldi \textit{et al.} in 1934/35, however, they did not ascribe it to a particular isotope \cite{Fer34,Ama35}.

\subsection*{$^{199}$Au}\vspace{-0.85cm}

$^{199}$Au was first observed by McMillan \textit{et al.} from the University of California at Berkeley in 1937 and reported in the paper \textit{Neutron-Induced Radioactivity of the Noble Metals} \cite{McM37}. Platinum was irradiated with slow neutrons and a 3.3 d activity was assigned to $^{199}$Au from the decay of $^{199}$Pt following chemical separation. ''...-the gold precipitate showed a 3.3-day activity. Reference to the isotope chart shows that one would expect Pt$^{199}$ to form unstable Au$^{199}$.'' This half-life agrees with the accepted value of 3.139(7) d.

\subsection*{$^{200}$Au}\vspace{-0.85cm}

$^{200}$Au was observed by Butement in 1951 at the Atomic Energy Research Establishment in Harwell, United Kingdom, as reported in \textit{New Radioactive Isotopes produced by Nuclear Photo-Disintegration} \cite{But51}. The isotope was produced by 23 MeV X-rays from a synchrotron in photonuclear reactions. ''The 48-minute gold has been made by Maurer and Ramm (1942) by the reactions Hg(n,p) and Tl(n,$\alpha$), thus proving the mass number to be wither 200 or 202. Its production also by Hg($\gamma$,p) shows that its mass number must be 200.'' This half-life agrees with the currently accepted values of 48.4(3) m. In addition to the quoted paper by Maurer and Ramm \cite{Mau42} the 48 m activity had also been previously observed by Sherr \textit{et al.} without a definite mass assignment \cite{She41}.

\subsection*{$^{201}$Au}\vspace{-0.85cm}

Butement and Shillito report the first observation of $^{201}$Au in Harwell, United Kingdom, in 1952 in their paper \textit{Radioactive Gold Isotopes} \cite{But52}. Fast neutrons produced from 20 MeV protons bombarding a beryllium target were used to irradiate mercury and thallium targets. Decay curves were measured for (n,p) (n,pn) on mercury and (n,$\alpha$) and (n,$\alpha$a) on thallium. The mass assignment was determined ``from the modes of production'' and was confirmed ``by studying the gold isotopes produced by ($\gamma$,p) reactions on enriched mercury isotopes.''  The half-life was determined via analysis of the decay curve to be 26~m. This value agrees with the currently accepted values of 26(1)~m. Butement had previously reported a 27 m activity with a probable assignment of$^{201}$Au or $^{203}$Au \cite{But50,But51}. Interestingly, Butement and Shillito do not acknowledge the identification of $^{200}$Au in reference \cite{But51}.

\subsection*{$^{202}$Au}\vspace{-0.85cm}

The paper \textit{New Isotope Au$^{204}$ and Decay of Au$^{202}$} by Ward \textit{et al.} in 1967 constitutes the first unambiguous identification of  $^{202}$Au \cite{War67}.  14.8 MeV neutrons, produced via the D(d,t) reaction from the University of Arkansas 400 KV Cockcroft-Walton linear accelerator bombarded natural mercury and thallium targets. Gamma- and beta spectra were recorded to identify $^{202}$Au. The extracted half-life of 30 s is consistent with the adapted half-life of 28.8(19) s. The authors do not consider their measurement a new discovery: ``It has been reported that a 25$\pm$5 sec activity is produced from the Hg$^{202}$(n,p)Au$^{202}$ and Tl$^{205}$(n,$\alpha$)Hg$^{202}$ reactions'' with a reference to the paper by Butement and Shillito \cite{But52}. However, Butement and Shillito do not claim the discovery because they were unable to distinguish between $^{202}$Au and $^{204}$Au: `` The 25-second activity is probably due to one of these isotopes'' \cite{But52}.

\subsection*{$^{203}$Au}\vspace{-0.85cm}

In the paper \textit{Radioactive Gold Isotopes} on the measurement of $^{200}$Au and $^{201}$Au Butement and Shillito also reported the first observation of $^{203}$Au in 1952 \cite{But52}. Fast neutrons produced from 20 MeV protons bombarding a beryllium target were used to irradiate mercury and thallium targets. Decay curves were measured for (n,p) (n,pn) on mercury and (n,$\alpha$) and (n,$\alpha$n) on thallium. The mass assignment of $^{203}$Au was determined ``from the modes of production'' and were confirmed ``by studying the gold isotopes produced by ($\gamma$,p) reactions on enriched mercury isotopes.''  The measured half-life of 55~s agrees with the value of 53(2)~s listed in the NUBASE evaluation \cite{Aud03}. It should be mentioned that the ENSDF database \cite{ENS08} quotes a half-life of 60(6)~s based on the measurement by Wennemann \textit{et al.} \cite{Wen94}. This paper entitled \textit{Investigation of new neutron-rich gold isotope $^{203}$AU and $^{205}$Au} does not acknowledge the work by Butement and Shillito.

\subsection*{$^{204}$Au}\vspace{-0.85cm}

A. Pakkanen \textit{et al.} identified $^{204}$Au for the first time at Jyv\"askyl\"a, Finland, in 1972: \textit{New $^{204}$Au Activity and the Decay of $^{202}$Au} \cite{Pak72}. Natural mercury targets were bombarded with 14-15 MeV neutrons from a Sames J150 neutron generator. Characteristic singles and coincidence $\gamma$-ray spectra were measured following the transport of the targets with a fast pneumatic transport system. ``The most probable assignment for the 40$\pm$3 s activity is $^{204}$Au, by the following arguments: ...'' The half-life agrees with the currently adapted value of 39.8(9) s. The previously reported value of 4(1) s \cite{War67} could not be confirmed. ``If the previously reported 4.0 s activity also belongs to $^{204}$Au, an explanation could be the two close-lying isomeric states.'' However, such a state has so far not been confirmed.

\subsection*{$^{205}$Au}\vspace{-0.85cm}

$^{205}$Au was discovered in 1994 by Wennemann \textit{et al.} at GSI, Darmstadt, Germany, reported in \textit{Investigation of New Neutron-Rich Gold Isotopes $^{203}$Au and $^{205}$Au} \cite{Wen94}. $^{205}$Au was produced in a deep-inelastic reaction from an 11.4 MeV/u $^{208}$Pb beam accelerated by the UNILAC accelerator impinging on a natural tungsten target. The isotope was identified by its $\beta$-decay properties: ``The half-life of $^{205}$Au of $T_{1/2}$ = 31(2)~s was obtained from the decay characteristics of the three strongest $\gamma$-rays of 379, 467, and 946 keV.'' This half-life is currently the only measured value for $^{205}$Au.

\section{Summary}
The discoveries of the known gold isotopes have been compiled and the methods of their production discussed. The limit for observing long lived isotopes beyond the proton dripline which can be measured by implantation decay studies has most likely been reached with the discovery of $^{170}$Au.
Only three gold isotopes ($^{189}$Au, $^{191}$Au and $^{204}$Au) were originally wrongly identified. The half-lives of four isotopes ($^{198}$Au, $^{200}$Au $^{201}$Au, and $^{202}$Au) were first measured without cleanly identifying the isotope. In the case of $^{198}$Au Fermi \textit{et al.} measured a decay curve following neutron capture on the only stable gold isotope ($^{197}$Au) \cite{Fer34,Ama35}. Although it is obvious that the activity resulted from the (n$\gamma$) reaction to $^{198}$Au we did not credit Fermi \textit{et al.} with the discovery of $^{198}$Au because they did not attribute the decay to this isotope. Assignment for the discovery of $^{179}$Au required a judgement call. While Demin \textit{et al.} \cite{Dem68} submitted their manuscript about four months earlier than Siivola \cite{Sii68}, we credit Siivola with the discovery, because Demin was aware of and had access to the (unpublished) results of Siivola.

\ack

This work was supported by the National Science Foundation under grants No. PHY06-06007 (NSCL) and PHY07-54541 (REU). MH was supported by NSF grant PHY05-55445. JQG acknowledges the support of the Professorial Assistantship Program of the Honors College at Michigan State University.

%%% Here we use thebibliography environment to produce the reference list,
%%% but you can use BibTeX as well:
%\bibliography{tmpadnd}

\newpage

\section*{EXPLANATION OF TABLE}\label{sec.eot}
\addcontentsline{toc}{section}{EXPLANATION OF TABLE}

\renewcommand{\arraystretch}{1.0}

\begin{tabular*}{0.95\textwidth}{@{}@{\extracolsep{\fill}}lp{5.5in}@{}}
\textbf{TABLE I.}
	& \textbf{Discovery of Gold Isotopes }\\
%\multicolumn{2}{p{0.95\textwidth}}{(Throughout this table,
%	italicized numbers refer to derived values)}
\\

Isotope &  Gold isotope \\
Author & First author of refereed publication \\
Journal & Journal of publication \\
Ref. &  Reference  \\
Method & Production method used in the discovery: \\
    & FE: fusion evaporation \\
    & LP: light-particle reactions (including neutrons) \\
    & MS: mass spectroscopy \\
    & SP: spallation \\
    & DI: deep-inelastic reactions \\
    & PN: photo-nuclear reactions \\
    & NC: neutron-capture reactions \\
Laboratory &  Laboratory where the experiment was performed\\
Country &  Country of laboratory\\
Year & Year of discovery  \\
\end{tabular*}
\label{tableI}

\newpage
\datatables

%% If the table is to span over the whole text width, we set:
\setlength{\LTleft}{0pt}
\setlength{\LTright}{0pt}

% To avoid ``Overfull \hboxes...'' decrease the intercolumn spacing:

\setlength{\tabcolsep}{0.5\tabcolsep}

\renewcommand{\arraystretch}{1.0}

%%\footnotesize % we need to squeeze the font size a lot!

\begin{longtable}[c]{%
@{}@{\extracolsep{\fill}}r@{\hspace{5\tabcolsep}} llllllll@{}}
\caption[Discovery of Gold Isotopes]%
{Discovery of Gold isotopes}\\[0pt]
\caption*{\small{See page \pageref{tableI} for Explanation of Tables}}\\
\hline
\\[100pt]
\multicolumn{8}{c}{\textit{This space intentionally left blank}}\\
\endfirsthead
Isotope & Author & Journal & Ref. & Method & Laboratory & Country & Year \\

$^{170}$Au & H. Mahmud & Eur. Phys. J. A & Mah02 & FE & Argonne & USA &2002 \\
$^{171}$Au & C.N. Davids & Phys. Rev. C & Dav97 & FE & Argonne & USA &1997 \\
$^{172}$Au & P.J. Sellin & Z. Phys. A & Sel93 & FE & Daresbury & UK &1993 \\
$^{173}$Au & J.R.H. Schneider & Z. Phys. A & Sch83 & FE & Darmstadt & Germany &1983 \\
$^{174}$Au & J.R.H. Schneider & Z. Phys. A & Sch83 & FE & Darmstadt & Germany &1983 \\
$^{175}$Au & C. Cabot & Nucl. Phys. A & Cab75 & FE & Orsay & France &1975 \\
$^{176}$Au & C. Cabot & Nucl. Phys. A & Cab75 & FE & Orsay & France &1975 \\
$^{177}$Au & A. Siivola & Nucl. Phys. A & Sii68 & FE & Berkeley & USA &1968 \\
$^{178}$Au & A. Siivola & Nucl. Phys. A & Sii68 & FE & Berkeley & USA &1968 \\
$^{179}$Au & A. Siivola & Nucl. Phys. A & Sii68 & FE & Berkeley & USA &1968 \\
$^{180}$Au & J.P. Husson & J. Phys.(Paris) Lett. & Hus77 & SP & CERN & Switzerland &1977 \\
$^{181}$Au & A. Siivola & Nucl. Phys. A & Sii68 & FE & Berkeley & USA &1968 \\
$^{182}$Au & P.G. Hansen & Nucl. Phys. A & Han70 & SP & CERN & Switzerland &1970 \\
$^{183}$Au & A. Siivola & Nucl. Phys. A & Sii68 & FE & Berkeley & USA &1968 \\
$^{184}$Au & P.G. Hansen & Phys. Lett. B & Han69 & SP & CERN & Switzerland &1969 \\
$^{185}$Au & G. Albouy & J. Phys. Radium & Alb60 & SP & Orsay & France &1960 \\
$^{186}$Au & G. Albouy & J. Phys. Radium & Alb60 & SP & Orsay & France &1960 \\
$^{187}$Au & W.G. Smith & Phys. Rev. & Smi55 & LP & Berkeley & USA &1955 \\
$^{188}$Au & W.G. Smith & Phys. Rev. & Smi55 & LP & Berkeley & USA &1955 \\
$^{189}$Au & N. Poffe & J. Phys. Radium & Pof60 & SP & Orsay & France &1955 \\
$^{190}$Au & G. Albouy & Compt. Rend. Acad. Sci. & Alb59 & SP & Orsay & France &1959 \\
$^{191}$Au & L.P. Gillon & Phys. Rev. & Gil54 & LP & Harvard & USA &1954 \\
$^{192}$Au & G. Wilkinson & Phys. Rev. & Wil48 & LP & Berkeley & USA &1948 \\
$^{193}$Au & G. Wilkinson & Phys. Rev. & Wil48 & LP & Berkeley & USA &1948 \\
$^{194}$Au & G. Wilkinson & Phys. Rev. & Wil48 & LP & Berkeley & USA &1948 \\
$^{195}$Au & G. Wilkinson & Phys. Rev. & Wil48 & LP & Berkeley & USA &1948 \\
$^{196}$Au & E. McMillan & Phys. Rev. & McM37 & LP & Berkeley & USA &1937 \\
$^{197}$Au & A.J. Dempster & Nature & Dem35 & MS & Chicago & USA &1935 \\
$^{198}$Au & M.L. Pool & Phys. Rev. & Poo37 & NC & Michigan & USA &1937 \\
$^{199}$Au & E. McMillan & Phys. Rev. & McM37 & NC & Berkeley & USA &1937 \\
$^{200}$Au & F.D.S. Butement & Proc. Phys. Soc. & But51 & PN & Harwell & UK &1951 \\
$^{201}$Au & F.D.S. Butement & Proc. Phys. Soc. & But52 & LP & Harwell & UK &1952 \\
$^{202}$Au & T.E. Ward & Phys. Rev. & War67 & LP & Arkansas & USA &1967 \\
$^{203}$Au & F.D.S. Butement & Proc. Phys. Soc. & But52 & LP & Harwell & UK &1952 \\
$^{204}$Au & T. Pakkanen & Nucl. Phys. A & Pak72 & LP & Jyv\"askyl\"a & Finland &1972 \\
$^{205}$Au & Ch. Wennemann & Z. Phys. A & Wen94 & DI & Darmstadt & Germany &1994 \\

\end{longtable}

\newpage

%% A long reference list can be squeezed using:
%\newcommand{\bibfont}{\footnotesize}

\normalsize

\begin{theDTbibliography}{1956He83}

\bibitem[Alb59]{Alb59t} G. Albouy, R. Bernas, M. Gusakow, N. Poffe, and J. Teillac, Compt. Rend. Acad. Sci. {\bf 249}, 407 (1959)
\bibitem[Alb60]{Alb60t} G. Albouy, M.M. Gusakow, and N.Poffe, J. Phys. Radium {\bf 21}, 751 (1960)
\bibitem[But51]{But51t} F.D.S. Butement, Proc. Phys. Soc. {\bf 64A}, 395 (1951)
\bibitem[But52]{But52t} F.D.S. Butement and R. Shillito, Proc. Phys. Soc. A {\bf 65}, 945 (1952)
\bibitem[Cab75]{Cab75t} C. Cabot, C. Deprun, H. Gauvin, B. Lagarde, Y. Le Beyec, and M. Lefort, Nucl. Phys. A {\bf 241}, 341 (1975)
\bibitem[Dav97]{Dav97t} C.N. Davids \textit{et al.}, Phys. Rev. C {\bf 55}, 2255 (1997)
\bibitem[Dem35]{Dem35t} A.J. Demster, Nature {\bf 136}, 65 (1935)
\bibitem[Gil54]{Gil54t} L.P. Gillon, K. Gopalakrishnan, and A. de-Shalit, Phys. Rev. {\bf 93}, 124 (1954)
\bibitem[Han69]{Han69t} P.G. Hansen \textit{et al.}, Phys. Lett. {\bf 28B}, 415 (1969)
\bibitem[Han70]{Han70t} P.G. Hansen, H.L. Nielsen, K. Wilsky, M. Alpsten, M. Finger, A. Lindahl, R.A. Naumann, and O.B. Nelson, Nucl. Phys. A {\bf 148}, 249 (1970)
\bibitem[Hus77]{Hus77t} J.P. Husson, C.F. Liang, and C. Richard-Serre, J. Phys.(Paris), Lett. {\bf 38}, L-245 (1977)
\bibitem[Mah02]{Mah02t} H. Mahmud \textit{et al.}, Eur. Phys. J. A {\bf 15}, 85 (2002)
\bibitem[McM37]{McM37t} E. McMillan, M. Kamen, and S. Ruben Phys. Rev. {\bf 52}, 375 (1937)
\bibitem[Pak72]{Pak72t} A. Pakkanen, T. Komppa, and H. Helppi, Nucl. Phys. A {\bf 184}, 157 (1972)
\bibitem[Pof60]{Pof60t} N. Poffe, G. Albouy, R. Bernas, M. Gusakow, M. Riou, and J. Teillac, J. Phys. Radium {\bf 21}, 343 (1960)
\bibitem[Poo37]{Poo37t} M.L. Pool, J.M. Cork, and R.L. Thornton, Phys. Rev. {\bf 52}, 239 (1937)
\bibitem[Sch83]{Sch83t} J.R.H. Schneider, S. Hoffman, F.P. He\ss berger, G. Munzenberg, W. Reisdorf, and P. Armbruster, Z. Phys. A {\bf 312}, 21 (1983)
\bibitem[Sel93]{Sel93t} P.J. Sellin, P.J. Woods, T. Davinson, N.J. Davis, A.N. James, K. Livingston, R.D. Page, and A.C. Shotter, Z. Phys. A {\bf 346}, 323 (1993)
\bibitem[Sii68]{Sii68t} A. Siivola, Nucl. Phys. A {\bf 109}, 231 (1968)
\bibitem[Smi55]{Smi55t} W.G. Smith and J.M. Hollander, Phys. Rev. {\bf 98}, 1258 (1955)
\bibitem[War67]{War67t} T.E. Ward, H. Ihochi, M. Karras, and J.L. Meason, Phys. Rev. {\bf 164}, 1545 (1967)
\bibitem[Wil48]{Wil48t} G. Wilkinson, Phys. Rev. {\bf 73}, 252 (1948)
\bibitem[Wen94]{Wen94t} Ch. Wennemann, W.-D. Schmidt-Ott, T. Hild, K. Krumbholz, V. Kunze, F. Meissner, H. Keller, R. Kirchner, and E. Roeckl, Z. Phys. A {\bf 347}, 185 (1994)

\end{theDTbibliography}

\end{document}